\documentclass[epj]{svjour}

\usepackage{amssymb}
\usepackage{amsmath}
\usepackage{graphicx}

\begin{document}

\title{$XXZ$-like phase in the F-AF anisotropic Heisenberg chain}

\author{Adolfo Avella \and Ferdinando Mancini \and Evgeny Plekhanov}

\institute{Dipartimento di Fisica ``E.R. Caianiello'' - Unit\`{a} CNISM di Salerno \protect\\
Universit\`{a} degli Studi di Salerno, I-84081 Baronissi (SA), Italy}

\abstract{
By means of the Density Matrix Renormalization Group technique, we have studied the region where $XXZ$-like behavior is most likely to emerge within the phase diagram of the F-AF anisotropic extended ($J-J'$) Heisenberg chain. We have analyzed, in great detail, the equal-time two-spin correlation functions, both in- and out-of- plane, as functions of the distance (and momentum). Then, we have extracted, through an accurate fitting procedure, the exponents of the asymptotic power-law decay of the spatial correlations. We have used the exact
solution of $XXZ$ model ($J'=0$) to benchmark our results, which clearly show the expected agreement. A critical value of $J'$ has been found where the relevant power-law decay exponent is independent of the in-plane nearest-neighbor coupling.}

\maketitle

\section{Introduction}\label{intro}

The spin-$1/2$ extended Heisenberg model is one of the most studied spin systems in the literature as it is the prototype of a geometrically frustrated quantum system and allows to study the interplay between two fundamental aspects of spin physics: quantization/fluctuation and frustration. Such competition is often responsible for the appearance of exotic phases and anomalous phenomena whose explanation is still under debate. The one-dimensional realization of the extended Heisenberg model (hereafter, $J$ stands for the nearest-neighbor exchange coupling and $J'$ for the next-nearest-neighbor one) has been mainly studied in the AF-AF regime ($J>0$, $J'>0$) according to the many experimental realizations. Only recently, the F-AF regime ($J<0$, $J'>0$) has come into the limelight as it has been proposed to explain the anomalous experimental behavior of compounds containing chains formed by edge-sharing $CuO_4$ plaquettes: $Rb_2Cu_2Mo_3O_{12}$ \cite{hase}, $NaCu_2O_2$ \cite{drechsler}, and $LiCuVO_4$ \cite{enderle}. In such class of compounds (both edge- and corner- sharing ones), the sign and the absolute value of $J$ depend sensitively on the $Cu$-$O$-$Cu$ bond angle and on the distance between copper and oxygen ions. In the compounds mentioned above, the $Cu$-$O$-$Cu$ bond angle is slightly larger than $90^\circ$. Accordingly, the usual antiferromagnetic nearest-neighbor exchange between $Cu$ ions becomes ferromagnetic (the $O_{2p}$ orbital hybridizing with the $3d$ orbital of one $Cu$ ion is almost orthogonal to that of the next $Cu$ ion). This nearly orthogonality makes the nearest-neighbor coupling in the edge-sharing case smaller, by more than an order of magnitude, than in the corner-sharing case. In edge-sharing compounds, the next-nearest-neighbor exchange coupling, through the $Cu$-$O$-$O$-$Cu$ path, is generally antiferromagnetic, usually with a magnitude of a few tens Kelvin. Despite its smallness, the next-nearest-neighbor exchange coupling interaction has a pronounced effect on the physical properties of these systems since the nearest-neighbor exchange coupling is also very small owing to the nearly orthogonality ($J'/J \approx -1/3$). On the theoretical level, it is known that for $J'/J > -1/4$, the ground state of the system is ferromagnetic \cite{bader}. At exactly $J'/J = -1/4$, a new singlet ground state, degenerate with the ferromagnetic one, appears and has a resonating-valence-bond form (under periodic boundary conditions: Ref.~\cite{hamada}). For $J'/J < -1/4$, only speculations are available, which range from a gapless singlet ground state \cite{white_03,allen} to a gapped one \cite{itoi}.

In this manuscript, we have studied the anisotropic version of the spin-$1/2$ extended Heisenberg chain with the intent of locating the region in its phase diagram where a $XXZ$-like behavior can be observed despite the presence of a finite next-nearest-neighbor coupling $J'$. In particular, we have analyzed the dependence of the critical exponents of the in-plane and out-of-plane correlation functions on $J'$. Quite unexpectedly, a critical value of $J'$ has been identified where the in-plane critical exponent does not depend on the in-plane nearest-neighbor coupling and the out-of-plane correlations qualitatively change their behavior.

\section{Definitions and Method} \label{model}

The anisotropic version of the F-AF Heisenberg chain reads as
\begin{multline}
H = -J_z \sum_i S^z_i S^z_{i+1} + J_\perp  \sum_i ( S^x_i S^x_{i+1} + S^y_i S^y_{i+1}) \\
+ J' \sum_i \mathbf{S}_i \mathbf{S}_{i+2}
\label{ham}
\end{multline}
where the nearest-neighbor out-of-plane coupling constant $J_z$ has been taken ferromagnetic ($J_z>0$) and the next-nearest-neighbor coupling constant $J'$ has been taken antiferromagnetic ($J'>0$). Results can be easily mapped between negative and positive signs of the nearest-neighbor in-plane coupling constant $J_\perp$ that has been fixed positive (antiferromagnetic). Hereafter, $J_z$ will be used as energy unit ($J_z=1$). The Hamiltonian model (\ref{ham}) opens up the possibility to analyze, in great detail, the role/relevance of the geometrical frustration in enhancing/quenching the quantum fluctuations. In fact, it contains both $J' \neq 0$, which can drive either gapless liquid \cite{white_03,allen} or gapped dimerized phases \cite{itoi}, and $J_z \neq J_\perp$, which can obviously drive quite different responses in- and out-of- plane. The very rich phase diagram, the competition between frustration and anisotropy and the absence of evident small parameters has attracted the attention of many researchers in the field. When $J_z$, $J_\perp$ and $J'$ are all of the same order of magnitude, exact (or quasi-exact) treatments, such as Bosonization and Renormalization Group, are inapplicable and numerical techniques are the only possible tool of investigation \cite{Plekhanov_01}. In this manuscript, we have used the Density Matrix Renormalization Group (DMRG) technique \cite{white_01,white_02} in order to obtain the ground state properties of the Hamiltonian (\ref{ham}). We have studied a $L=100$-site chain, subject to open-boundary conditions, and we have retained up to the lowest $200$ eigenstates of the reduced density matrix in the basis of each DMRG block. The real-space spin-spin correlation functions [both in-plane ($\langle S^{x,y}_r S^{x,y}_{r'} \rangle$) and out-of-plane ($\langle S^{z}_r S^{z}_{r'} \rangle$)] have been calculated, as functions of the distance $d=|r-r'|$, by choosing $r$ and $r'$ symmetrically with respect to the center of the system. The momentum-space correlation functions have been obtained as the Fourier transform of the real-space ones. It is worth noting that Fourier transform is generally useless when open-boundary conditions are used: functions in direct space generally depend on two set of coordinates and the transformed functions generally depends on two momenta. However, larger is the system under analysis, smaller are the non-diagonal terms in the transformed function (they are exactly zero when periodic boundary conditions are used). We have carefully checked that our data satisfy this condition: non-diagonal terms in momentum space are safely negligible in the whole region of parameter space under analysis. Such test has confirmed that $100$ sites are sufficient to our purposes and that the effects of open boundary conditions are under control.

\begin{figure}[t!]
\includegraphics[width=0.49\textwidth]{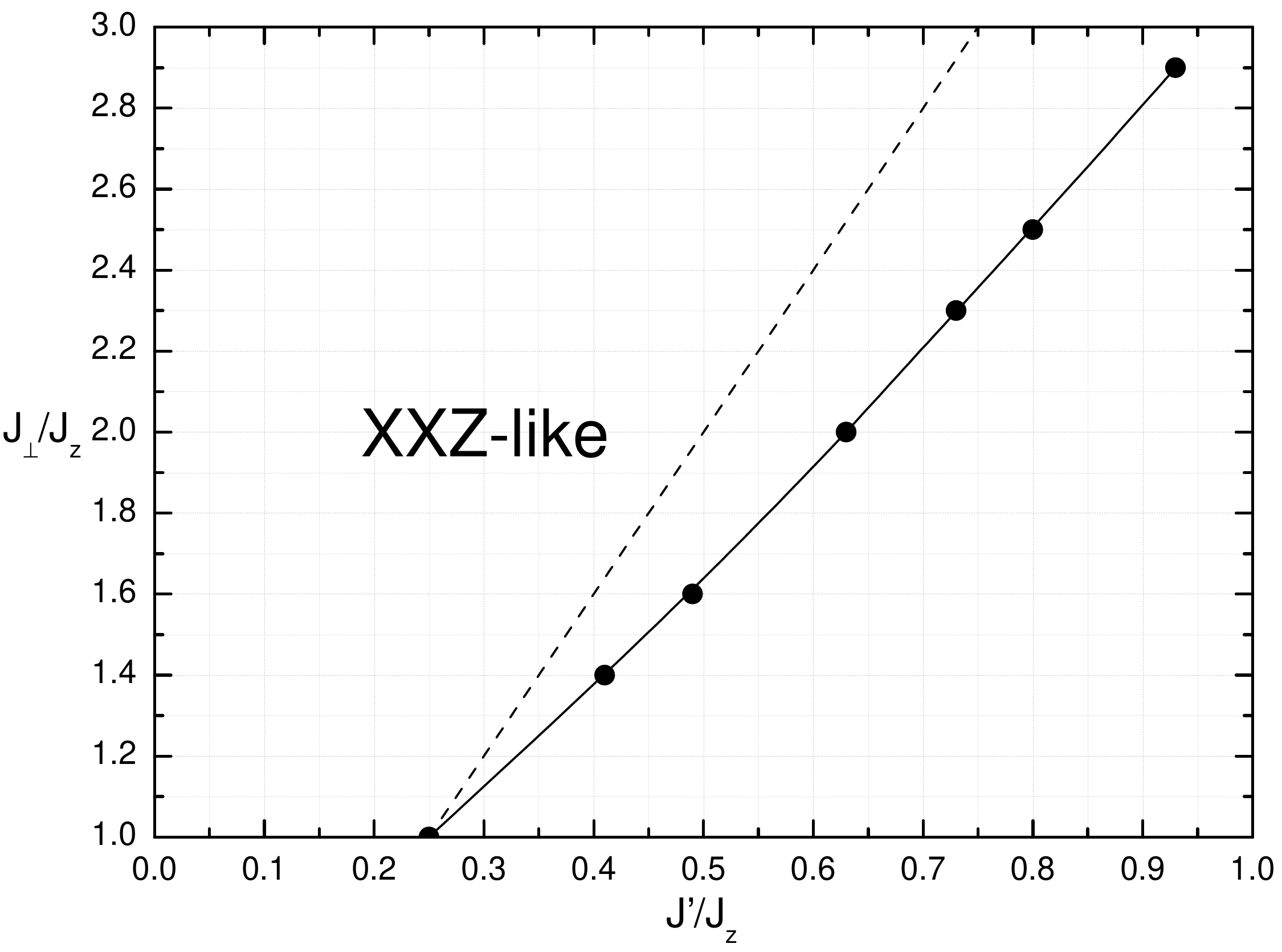}
\caption{$XXZ$-like region in the ($J'$, $J_\perp$) plane. The solid line is a guide to the eye. The dashed line is the classical transition line between an in-plane antiferromagnet and an in-plane incommensurate spiral phase.}
\label{fig0}
\end{figure}

\section{Results}

In the region under analysis ($J_\perp/J_z>1$, $J'/J_z<1$), the classical phase diagram (i.e. the one obtained for classical vectors of constant length) presents a transition between an in-plane antiferromagnet (main momentum is $k=\pi$) and an in-plane incommensurate spiral phase with main momenta $k=\pi\pm\arccos(J_\perp/4J')$ for $J'>J_\perp/4$ (dashed line in Fig.~\ref{fig0}). Quantum fluctuations modify this picture. Quasi-long-range correlations, with power-law decay at large distances, replace the long-range correlations with constant values. In the simple $J'=0$ case ($XXZ$ model), as soon as the rotational invariance is broken, the long-distance correlation functions assume a simple power-law form \cite{giamarchi}:
\begin{equation}
\begin{split}
&\langle S^{x,y}_0 S^{x,y}_r \rangle \xrightarrow{r \to \infty} (-1)^r A_\perp r^\eta + B_\perp r^{\eta + 1/\eta}\\
&\langle S^{z}_0 S^{z}_r \rangle \xrightarrow{r \to \infty} (-1)^r A_z r^{1/\eta} + B_z r^{-2}
\end{split}
\label{giamarchi}
\end{equation}
where $\eta=-\arccos(J_z/J_\perp)/\pi$, $1 < J_\perp/J_z$, $A_\perp$ has been determined in Ref.~\cite{lukyanov}.

\begin{figure*}[t!]
\includegraphics[width=0.49\textwidth]{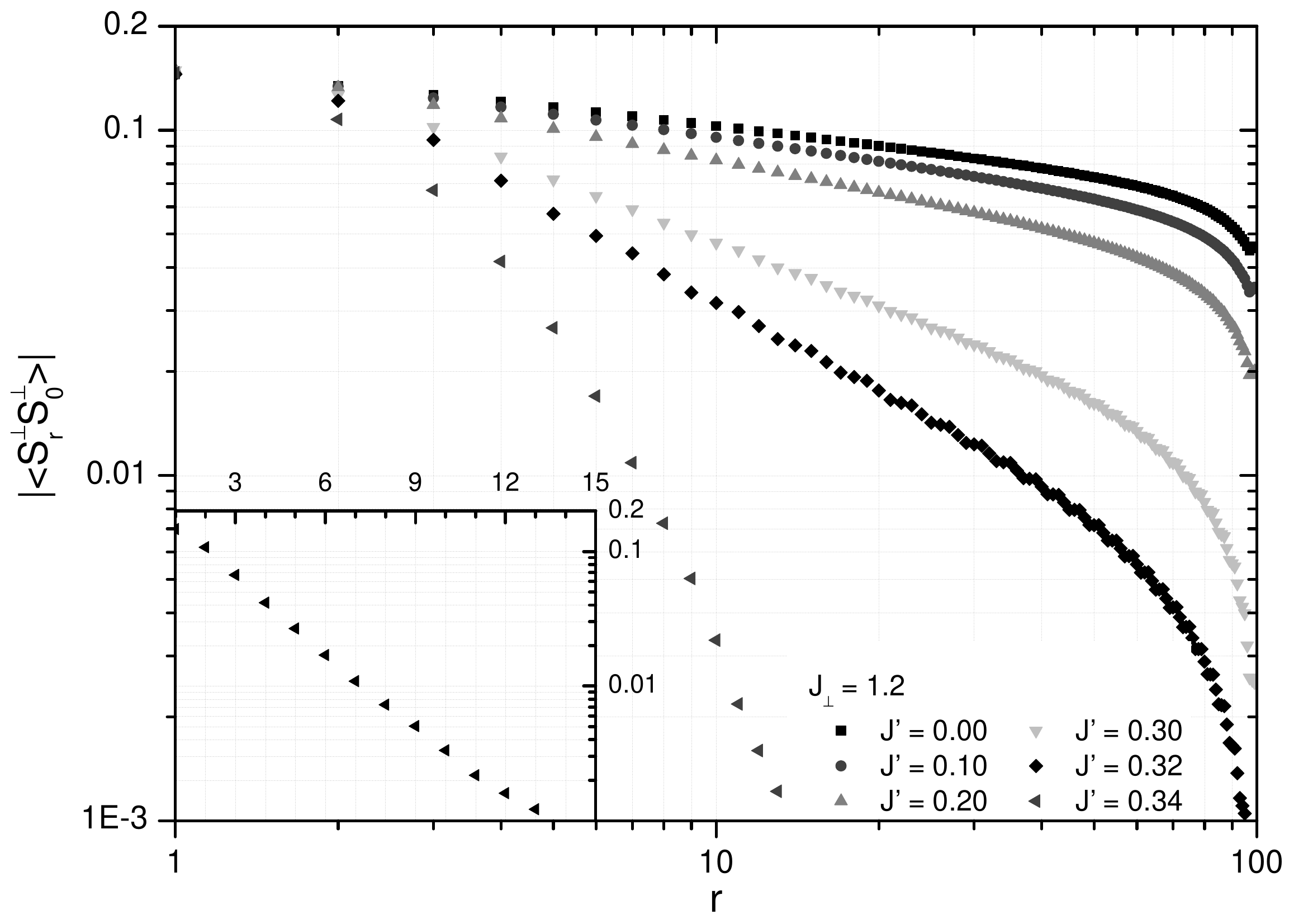}
\includegraphics[width=0.49\textwidth]{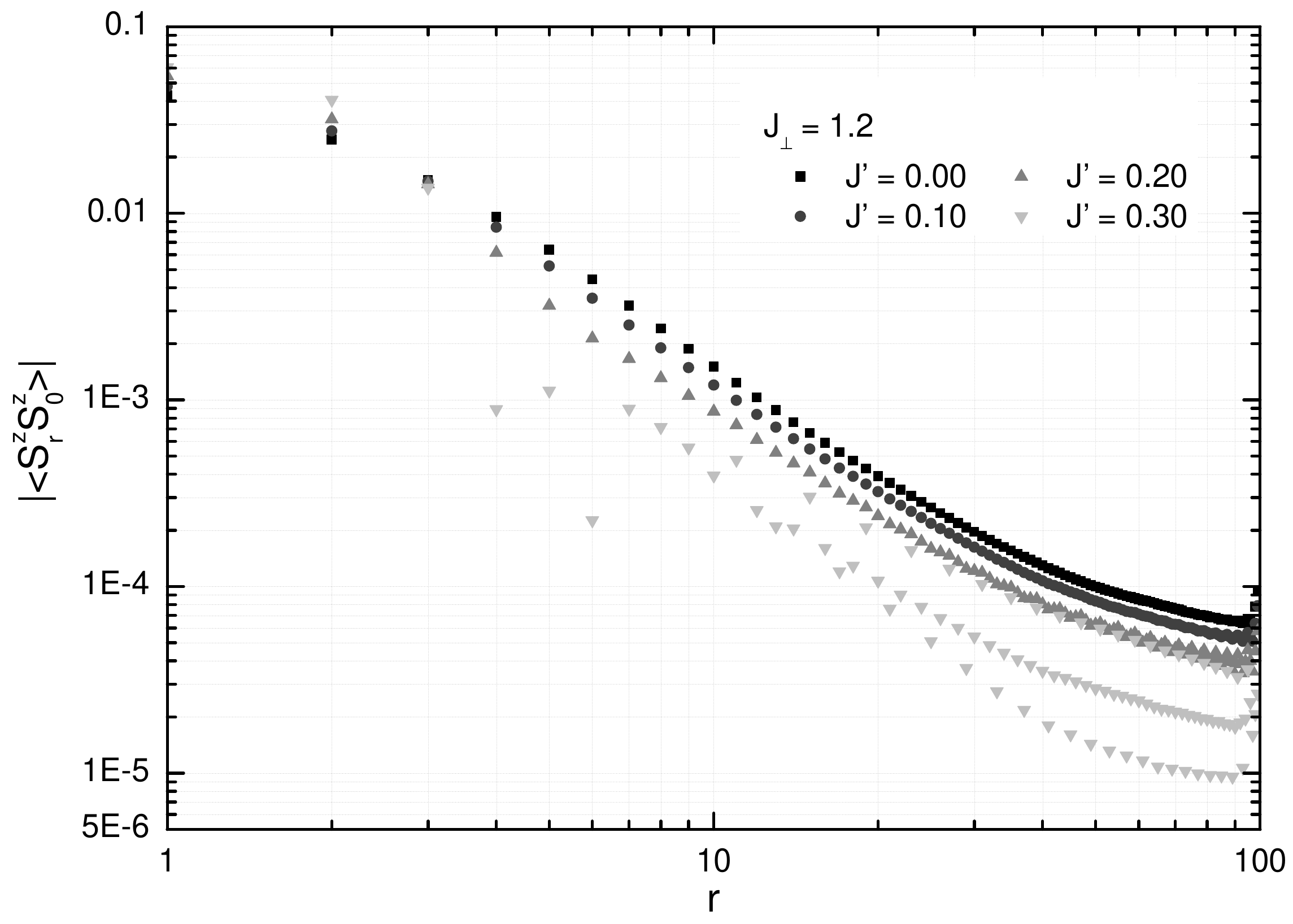}
\includegraphics[width=0.49\textwidth]{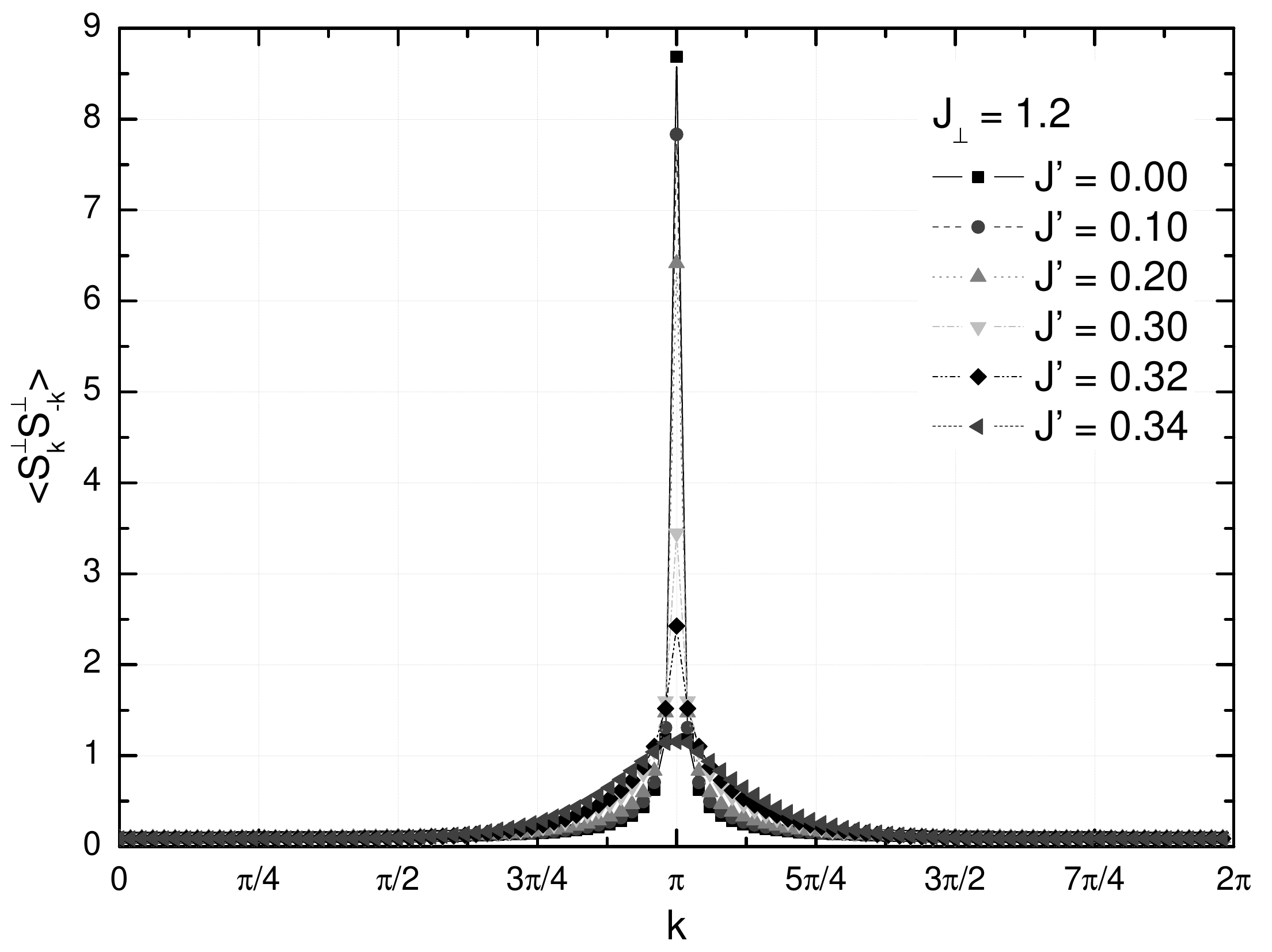}
\includegraphics[width=0.49\textwidth]{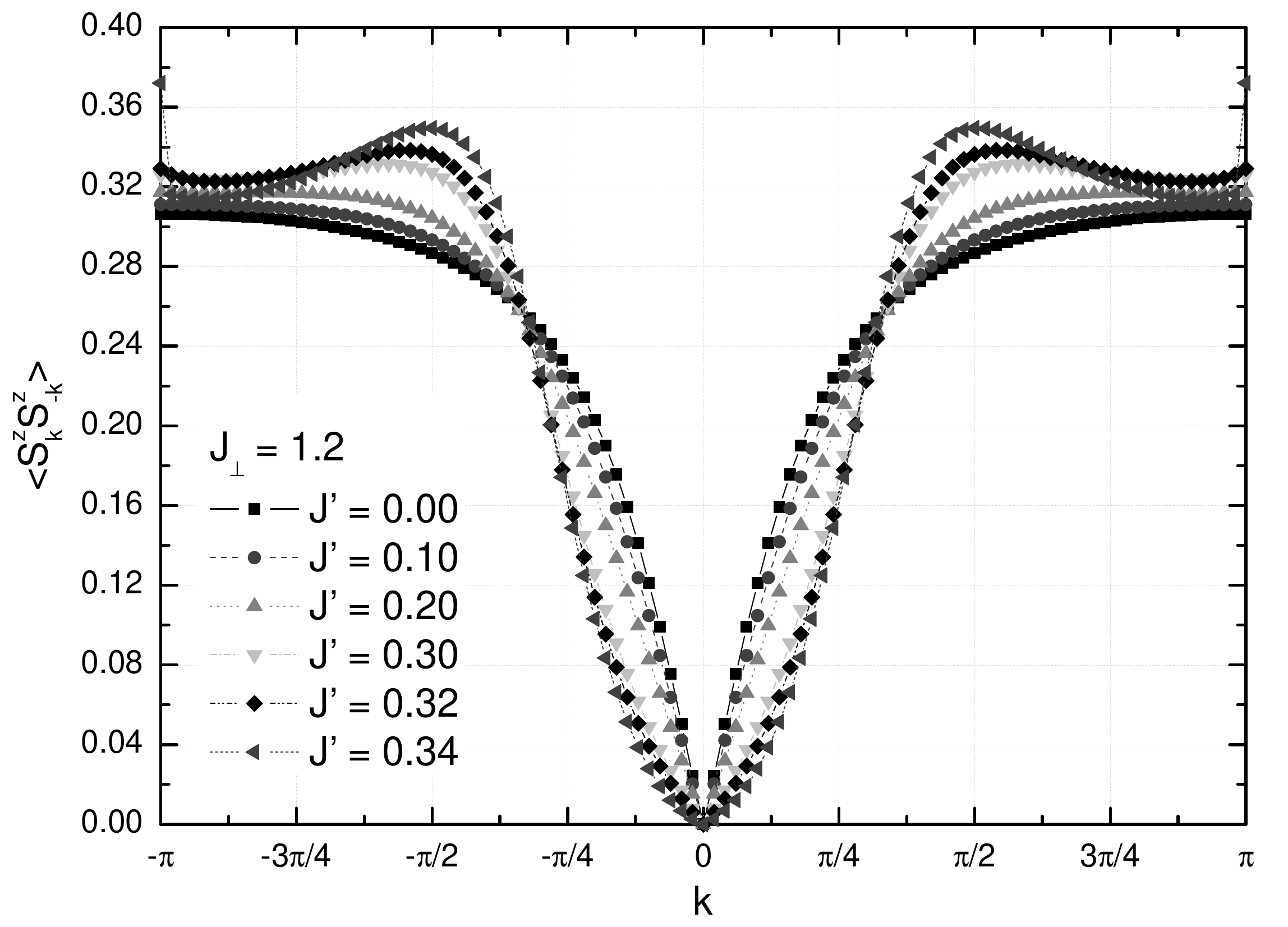}
\caption{Equal-time two-spin correlation functions in-plane $\langle S^\perp S^\perp \rangle$ (left panels) and out-of-plane $\langle S^z S^z \rangle$ (right panels) as functions of the distance $r$ (top panels) and momentum $k$ (bottom panels) for $J_\perp=1.2$ and different values of $J'$.}
\label{fig1}
\end{figure*}

We have analyzed the equal-time two-spin correlation functions, both in-plane $\langle S^\perp_0 S^\perp_r \rangle$ and out-of-plane $\langle S^{z}_0 S^{z}_r \rangle$, as functions of the distance $r$, together with their Fourier transforms, in order to easily recognize short-, quasi-long- and long- range orders and their characteristic features. A long-range order is characterized by a correlation in momentum space showing a Dirac delta (a peak of height proportional to the size of the system) at some momentum $\bar{k}$, which corresponds to a correlation in direct space with constant amplitude and modulation $\mathrm{e}^{\mathrm{i}\bar{k}r}$. The system under analysis shows such a behavior only in the ferromagnetic phase ($J_\perp /J_z<1$, $J'/J_z \lesssim 0.25$). A quasi-long-range order is characterized by a correlation in momentum space showing a power-law divergence $1/\left|k-\bar{k}\right|^\eta$ (a peak on a finite system) with exponent $0 \leqslant \eta < 1$ at some momentum $\bar{k}$. Such a behavior corresponds to a correlation in direct space showing a power-law decay $1/\left|r\right|^{1-\eta}$ with exponent $0 < 1-\eta \leqslant 1$ and modulation $\mathrm{e}^{\mathrm{i}\bar{k}r}$. We have extracted $\bar{k}$ from the correlations functions in momentum space and $\eta$ from those in direct space. In fact, the relation $1/\left|k\right|^\eta \to 1/\left|r\right|^{1-\eta}$ holds only for infinite systems and is badly violated for finite systems. A short-range order is characterized by a correlation in momentum space showing a Laurentian at some momentum $\bar{k}$ of width $\Delta$, which corresponds to a correlation in direct space showing an exponential decay $\mathrm{e}^{-\Delta r}$ with characteristic length $1/\Delta$ and modulation $\mathrm{e}^{\mathrm{i}\bar{k}r}$.

\begin{figure*}[p]
\includegraphics[width=0.49\textwidth]{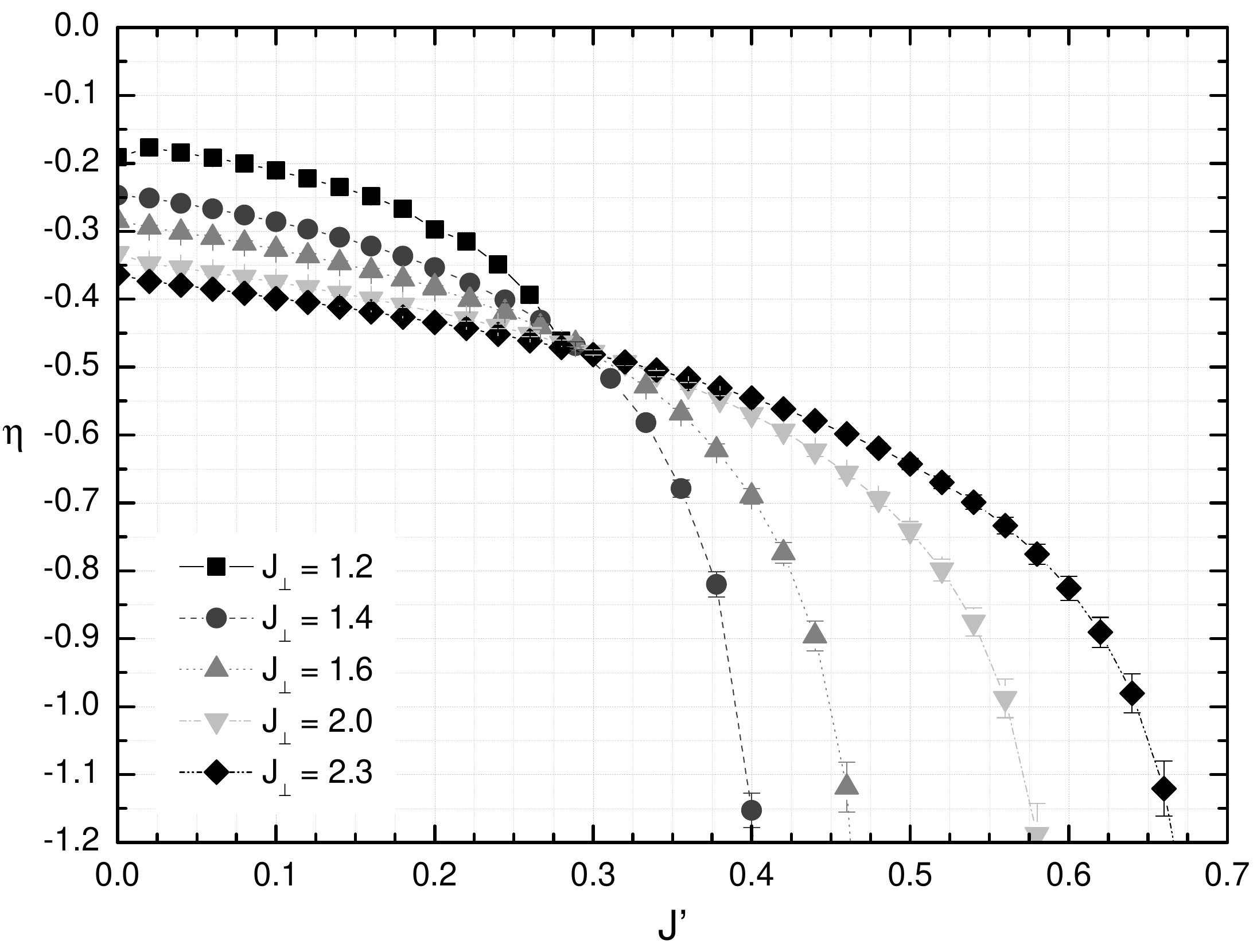}
\includegraphics[width=0.49\textwidth]{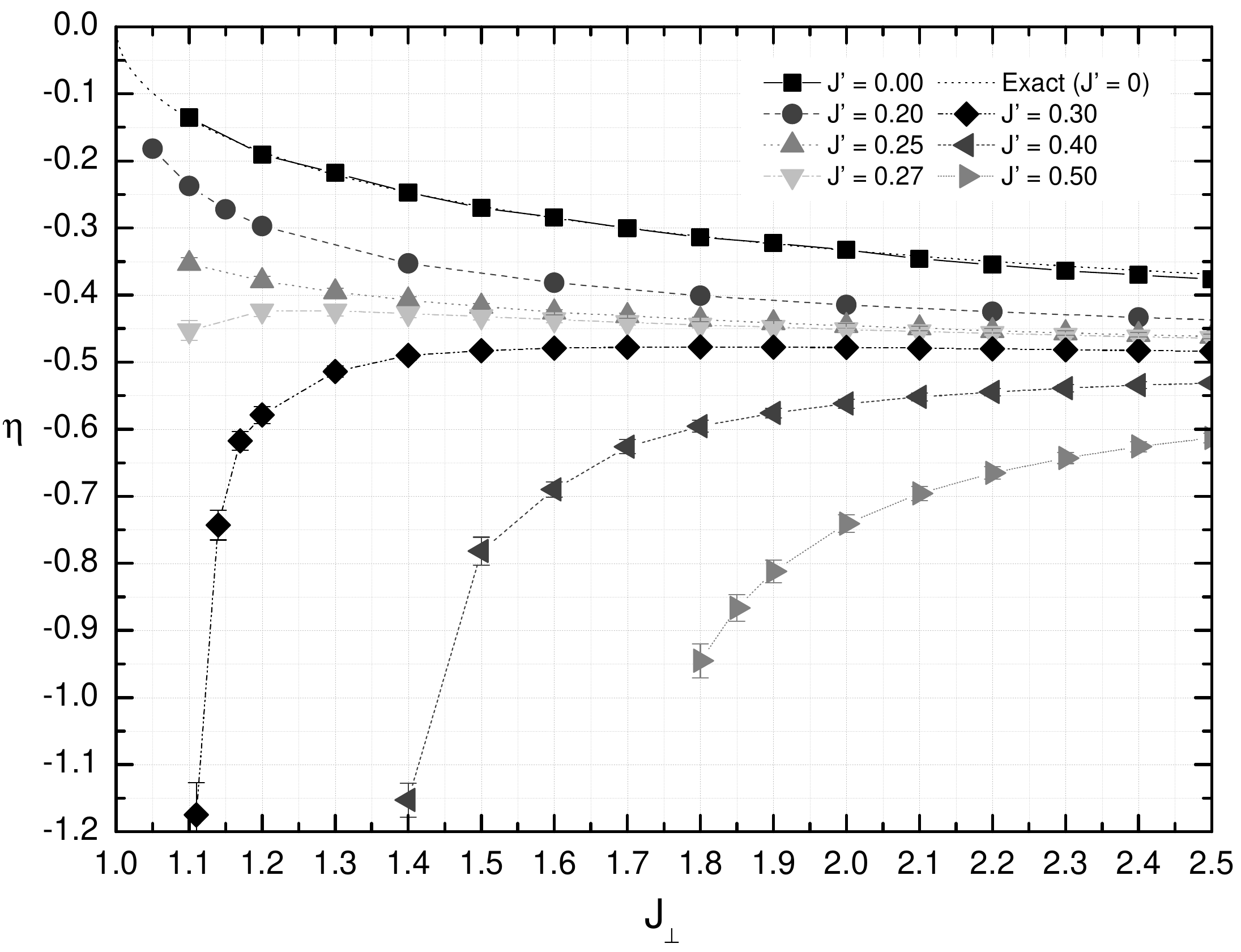}
\caption{Exponent $\eta$ of the power-law decay over distance of $\langle S^\perp S^\perp \rangle$ for different values of $J_\perp$ (left panel) and $J'$ (right panel).}
\label{fig2}
\end{figure*}

\begin{figure*}[p]
\includegraphics[width=0.49\textwidth]{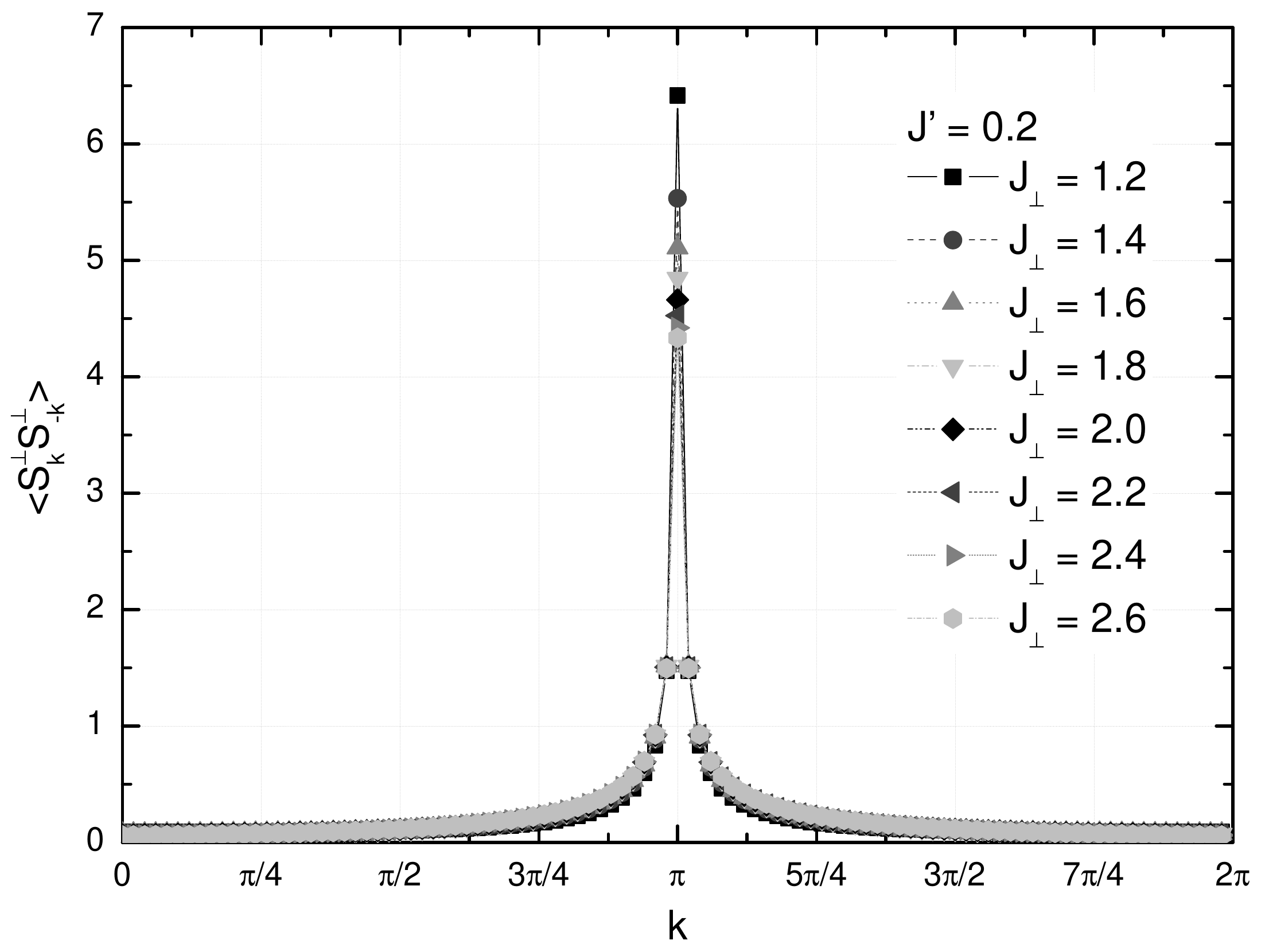}
\includegraphics[width=0.49\textwidth]{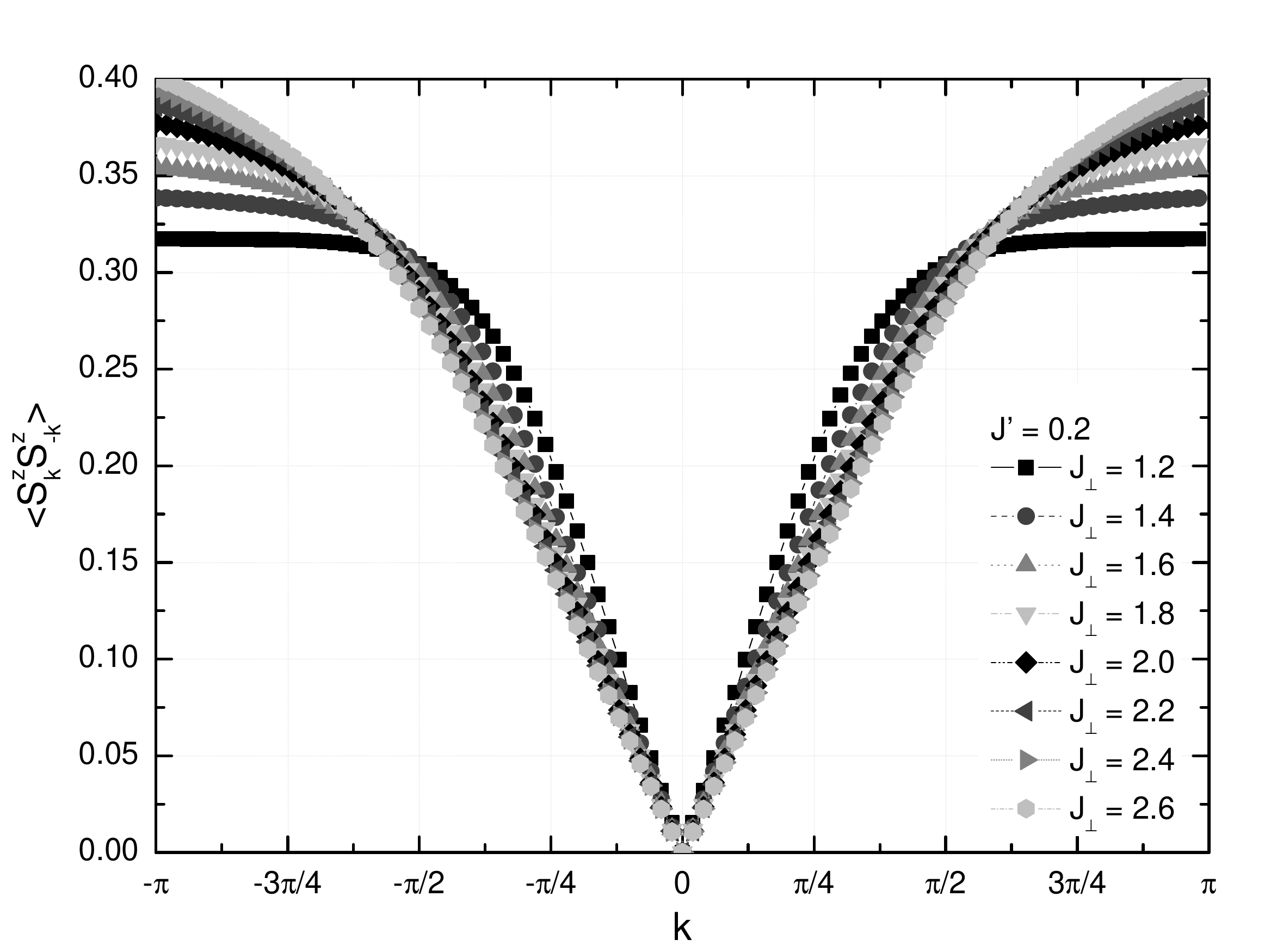}
\includegraphics[width=0.49\textwidth]{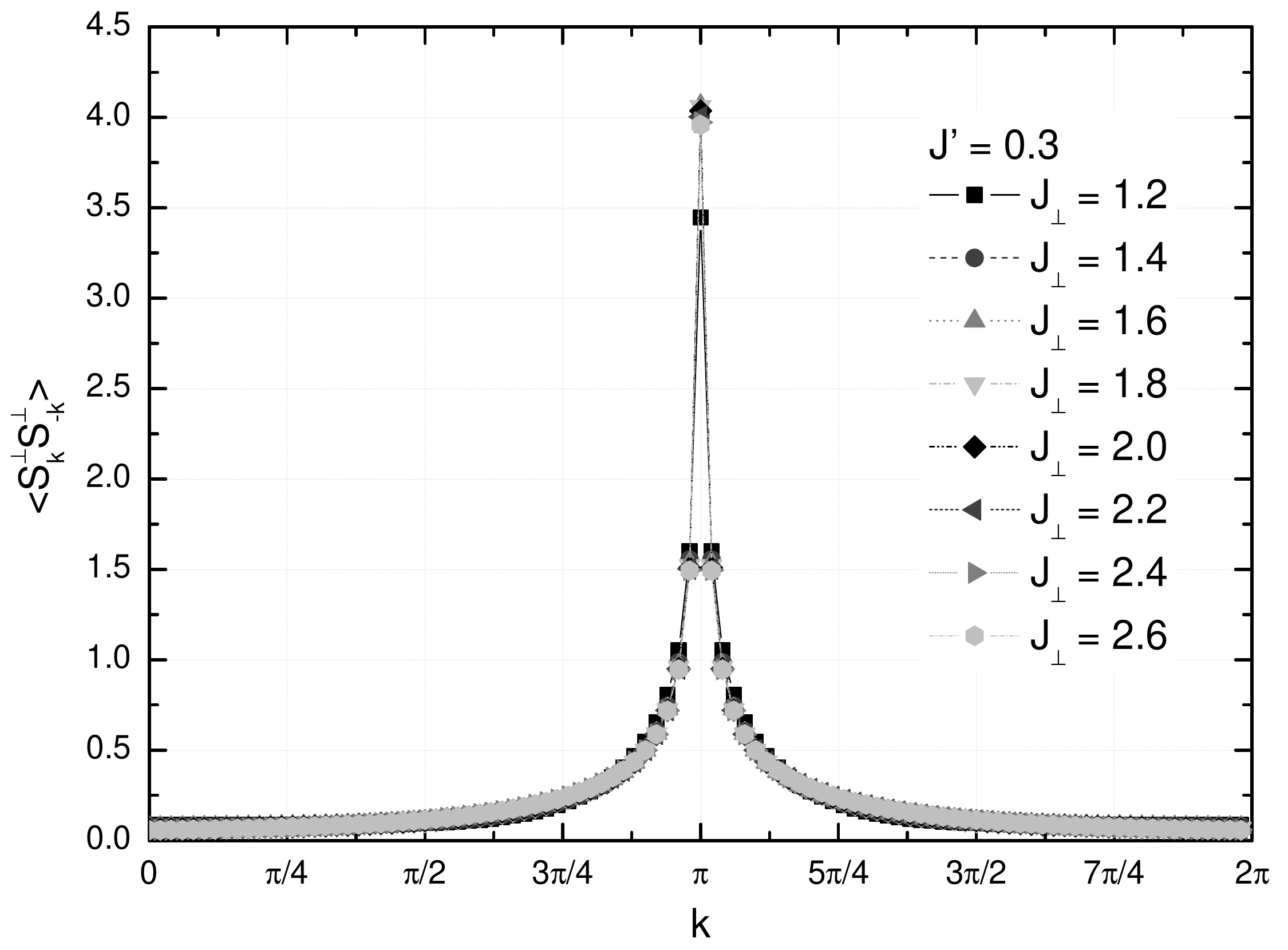}
\includegraphics[width=0.49\textwidth]{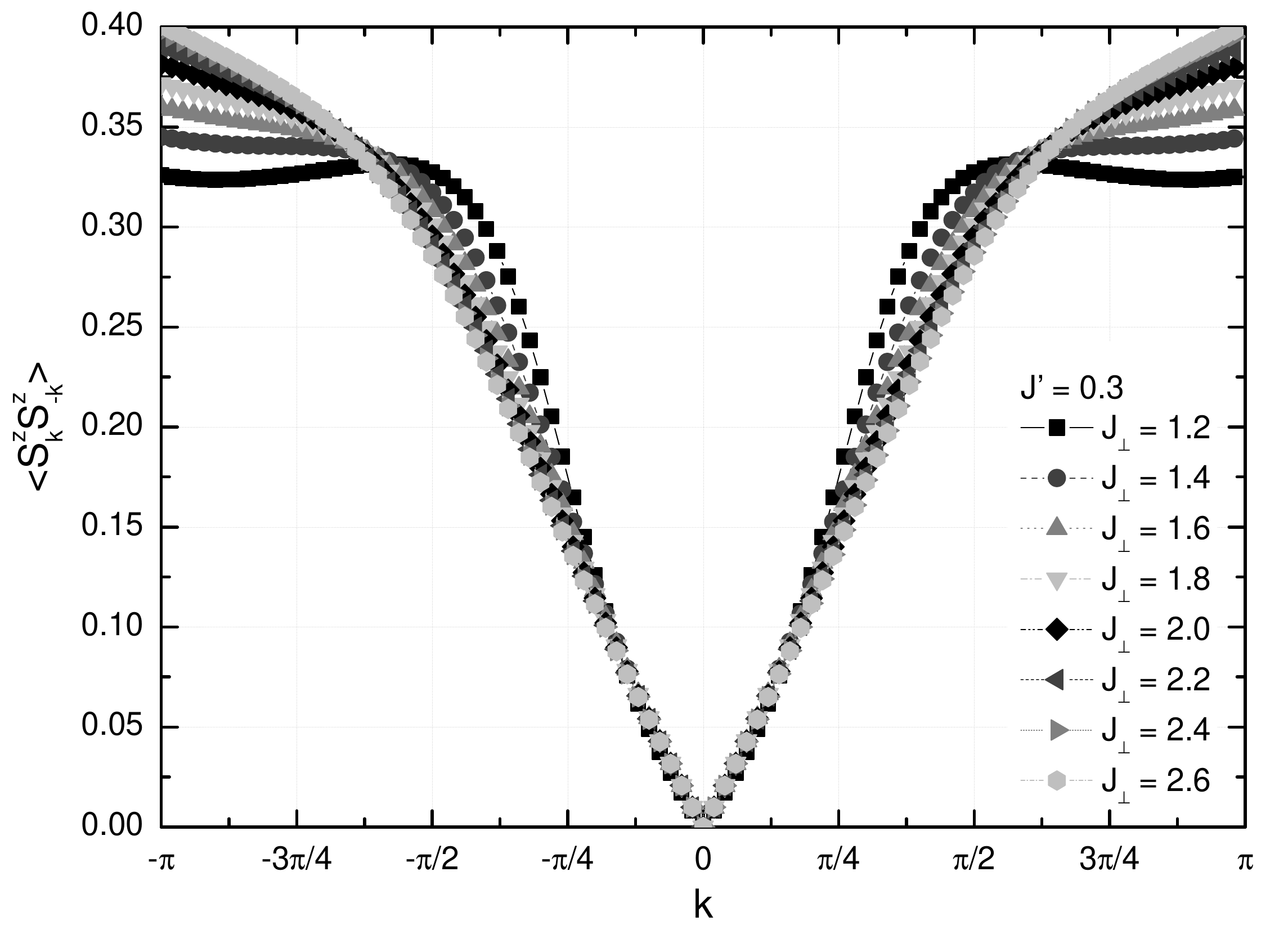}
\caption{Equal-time two-spin correlation functions in-plane $\langle S^\perp S^\perp \rangle$ (left panels) and out-of-plane $\langle S^{z} S^{z} \rangle$ (right panels) as functions of the momentum $k$ for $J'=0.2$ and $J'=0.3$.}
\label{fig3}
\end{figure*}

In Fig.~\ref{fig1}, we have reported the equal-time two-spin correlation functions, both in-plane $\langle S^\perp S^\perp \rangle$ (left panels) and out-of-plane $\langle S^{z} S^{z} \rangle$ (right panels), as functions of the distance $r$ (top panels) and momentum $k$ (bottom panels) for $J_\perp=1.2$ and different values of $J'$; results differ only quantitatively for higher values of $J_\perp$. Up to $J'=0.32$, the in-plane correlations clearly show a simple quasi-long-range (linear in log-log scale) antiferromagnetic order (left panels): a well defined peak centered at $k=\pi$ is clearly visible. At $J'=0.34$, the in-plane correlations lose the quasi-long-range character: the decay is exponential (top left panel and its inset: linear in semi-log scale) and the peak at $k=\pi$ becomes a Laurentian (bottom left panel). The out-of-plane correlations (right panels), completely absent in the classical case, clearly show a linear well ($|k| \to r^{-2}$) at $J'=0$ that changes its concavity on increasing $J'$ (bottom right panel). A small peak develops at $k=\pm\pi$ for $0.2 \leqslant J' < 0.34$. For larger values of $J'$, the out-of-plane correlations develop a small (short-range --- linear in semi-log scale) antiferromagnetic component. If we use the appearance of an exponential decay in the in-plane correlation functions as the signature of a transition to a different phase, we can locate the $XXZ$-like region in the ($J'$, $J_\perp$) plane and draw the transition line shown in Fig.~\ref{fig0}) (circles and solid line). It is worth noting that, for the largest reported values of $J'$ ($J' \geqslant 0.30$), the results for the out-of-plane correlations in direct space fall on three different curves (top right panel) and, correspondingly, in momentum space (bottom right panel), it is clearly visible the emergence of peaks in proximity of $k=\pm\pi/2$. Such behavior is most probably connected with a spiral phase, reminiscent someway of the classical one, appearing beyond the the massive phase that we are using to bound the XXZ-like phase under analysis. We will not discuss further neither the massive nor the spiral phase as they are out of the focus of this manuscript.

It is worth noticing that the observed behaviors are quite well described by (\ref{giamarchi}). Then, if we do believe that the (\ref{giamarchi}) do not change for small enough values of $J'$, our analysis simply accounts for the dependence of the critical exponent $\eta$ on the value of $J'$. While the in-plane correlations have as leading term the antiferromagnetic one for any finite value of $\eta$, the out-of-plane correlations present a competition between the two components (antiferromagnetic and ferromagnetic) depending on the value of $\eta$. In particular, for values of $\eta$ larger than $-1/2$ ($-1/2< \eta \leqslant 0$), the ferromagnetic component is dominant, while the antiferromagnetic one prevails for $\eta < -1/2$. Now, for $J'=0$, $\eta$ can only decrease from $0$ to $-1/2$ on increasing $J_\perp$. Accordingly, the out-of-plane correlations can acquire an antiferromagnetic component that has an overall exponent ($1/\eta \leqslant -2$), at most, as small as the ferromagnetic one ($-2$). For $J'>0$, $\eta$ can reach (see Fig.~\ref{fig2}), for a fixed value of $J_\perp$, much lower values than $-1/2$ and, in principle, it becomes possible to explore cases in which, for the out-of-plane correlations, the antiferromagnetic component is dominant.

In Fig.~\ref{fig2}, we have reported the critical exponent $\eta$ of the power-law decay of the in-plane correlation functions for different values of $J_\perp$ (left panel) and $J'$ (right panel). In the relevant parameter region, the exponent $\eta$ has been extracted through an accurate fitting procedure of the in-plane correlation functions in direct space (in log-log scale plots). In order to benchmark such a fitting procedure, we have also analyzed the exactly solvable $J'=0$ case and compared our results with the exact ones, resulting in an excellent agreement (right panel). It is clearly visible that there exists a critical value $J'_c \lessapprox 0.3$ where $\eta$ is independent of $J_\perp$. It is worth noting that for $J'=J'_c$, $\eta$ is very close to $-1/2$, the critical value for the out-of-plane correlations (ferromagnetic vs. antiferromagnetic).

In Fig.~\ref{fig3}, we have reported the behavior of both in-plane (left panels) and out-of-plane (right panels) correlation functions in momentum space for values of $J'$ below ($J'=0.2$) and above ($J'=0.3$) the critical one in order to extract any significant difference in the behavior across such a transition. Below $J'_c$, the in-plane correlation functions show a monotonic decrease of the peak intensity on increasing $J_\perp $. Below $J'_c$, the peak generally has a lower intensity with respect to the former case. Moreover, and more interestingly, the intensity of the peak first increases, on increasing $J_\perp $, and then decreases and rapidly saturates to values similar to those obtained below $J'_c$. The out-of-plane correlation functions simply show, at first glance, a systematic increase of their value at $\pm\pi$ on increasing $J_\perp$, for both values of $J'$. What is really reminiscent of the behavior found for $\langle S^\perp S^\perp \rangle$ is instead the slope of $\langle S^{z} S^{z} \rangle$ at $\pm\pi$. This latter changes sign at exactly the same value where the maximum of the intensity is reached for $\langle S^\perp S^\perp \rangle$. Moreover, for those values of $J_\perp$ where the intensity of the peak in $\langle S^\perp S^\perp \rangle$ increases, two well-defined shoulders appear in $\langle S^{z} S^{z} \rangle$.

\section{Conclusions}

In conclusions, we have found a region in the phase diagram of the spin-$1/2$ extended anisotropic Heisenberg chain where a $XXZ$-like behavior persists. We have extracted the exponent of the power-law decay and found a critical value of $J'$ where the former is practically independent of $J_\perp$.

\bibliographystyle{epj}
\bibliography{paper}

\end{document}